# Bi-Mode Foster's Equivalent Circuit of Arbitrary Planar Periodic Structures and its Application to Design Polarization Controller Devices

Gerardo Perez-Palomino, and Juan. E. Page

*Abstract—* **A Foster's equivalent circuit for 2-D Planar Periodic Structures (PPSs) that exhibit an arbitrary geometry is presented for first time in this paper. The proposed 4-port network shows an invariant circuit topology to the PPS geometry and is completely comprised of invariant-frequency lumped elements. The circuit is the simplest in terms of number of elements within the bi-mode bandwidth for a certain geometry, and its topology makes it possible both a standardized process to obtain the equivalent circuit of an arbitrary geometry and the use of the circuit theory to design a multitude of devices. To perform a validation, the equivalent circuit of two different PPSs (a rotated dipole and a defected slotted ring) have been obtained and analyzed in single and multi-layer configurations. The circuit is also used as a tool to develop Elliptical to Linear (or Circular) Polarization converters. One of the designs presented at 20 GHz (in transmission) converts an incident elliptical polarization of Axial Ratio 5 and tilted φ=20º into a purely linear vertical polarization of XP<-30 dB, with near-zero reflections and insertion loss better than 0.1 dB. The designed device is also manufactured and tested, and the measurements are in good agreement with simulations.**

## I. INTRODUCTION

THE 2-D periodic distributions of planar metallic scatters, known as Planar Periodic Structures (PPSs), are commonly used in the development of a multitude of devices that operate in non-guided systems. Spatial and spectral filters (known as Frequency Selective Surfaces, FSS [1]-[5]), Polarization control devices such as Linear to Circular Polarization (LP-CP) Converters [6]-[13] or Polarization Rotators [14]-[16], and aperture antennas (reflectarrays, transmitarrays [17]-[21]), are examples that use this type of planar configuration mainly because of its simplicity to be manufactured. Some of these works use equivalent circuits to make the design process much more efficient than when using electromagnetic (EM) simulators, especially to find an appropriate solution (pre-design) that allows performing an EM-optimization of problems that involve thousands of elements with guarantee of success.

Significant efforts have been invested in the literature to obtain accurate equivalent circuits of a variety of PPS geometries [22]-[35]. The strategies used to obtain a suitable equivalent circuit for one or multiple modes are usually based in the previous knowledge of the physical behavior of the structure, and usually provide phenomenological models that describe every physical interaction of the fields in the particular structure of the cell. Thus, the topologies of the resulting equivalent circuits are dependent on the geometry of the PPS, and comprise a relatively high number of elements that can be frequency-dependent, especially when dielectrics different from the vacuum are used. Thus, the complexity of these equivalent circuits may be problematic from the point of view of the design.

The main parameters to evaluate the viability of an equivalent circuit to describe a certain structure are: the bandwidth within the circuit can accurately predict the electrical behavior of the structure, the number of elements used, the topology of the circuit and the type of elements used (dependent or independent on the frequency). Another important parameter is the type of excitation and the number of modes that the circuit is able to consider, as some circuits of PPSs are proposed just for normal incidence or for oblique incidence in the main planes.

In this regard, the Foster's circuits present certain properties that make them interesting as an additional tool for the designer together with the aforementioned circuits reported in the state of the art. On the one hand, a Foster's form does not represent every phenomenological energetic interchange in the structure, but they are encompassed in lumped elements that are frequency-independent. Thus, the resulting circuit is the simplest in terms of number of elements [36]. On the other hand, a Foster's circuit presents a fixed topology (T, Pi, lattice, etc), which have been widely studied in the classic circuit

Manuscript received November 25, 2019. This work was supported in part by the Spanish Ministry of Economy, Industry and Competitiveness, under the project TEC2016-75103-C2-1-R.

G. Perez-Palomino and J. E. Page, are with the Group of Applied Electromagnetics (GEA), Universidad Politécnica de Madrid, E-28040, Madrid, Spain (e-mail: gerardo.perezp@upm.es).

theory in a multitude of design strategies.

Some works have proposed and used Foster's circuits to model and design devices based on waveguides [37]-[38] and PPSs. However, these circuits in the case of PPSs have been limited to symmetrical geometries with respect to the periods, which provide two independent circuits for each mode of excitation (TE or TM) (i.e. [6], [19]). In a previous work [12], we proposed a bi-mode (4-ports) Foster's equivalent circuit to consider the interaction of both modes of excitation for normal incidence at the same time. The circuit proposed in [12] allows the "anisotropic" geometries that are necessary to perform polarization converters to be considered, so it was used to obtain a systematic procedure to develop LP-CP converters made up an arbitrary number of layers using the theory of 4-port networks. However, the circuit in [12] is valid just in case that the PPS geometry exhibits dual diagonal symmetry, as the circuit was obtained from an even-odd decomposition approach.

In this paper a bi-mode Foster's equivalent circuit of 2-D PPSs with no restrictions on the symmetry of the geometry is presented. The approach used to obtain this circuit is completely different from that used in [12], and is based in the continuity of the field. The proposed 4-port network shows an invariant circuit topology to the PPS geometry, and is completely made up of invariant-frequency lumped elements independently of the medium used. The circuit is the simplest in terms of number of elements within the bi-mode bandwidth of the PSS cells for a certain geometry, and its topology makes it possible both an standardized process to obtain the equivalent circuit of an arbitrary geometry (for which the number of lumped elements is the only variable that depends on the geometry), and the use of the classic circuit theory to design a multitude of devices. Once the fundamentals and the circuit are presented in section II, the proposed equivalent circuit is obtained and validated for two different non-symmetrical PPS geometries in single and multi-layer configuration (section III). Finally, the circuit is also used in section IV as a pre-design tool to develop Elliptical to Linear (or Circular) Polarization converters (EP-LP or EP-CP). One of the transmissive EP-LP designed converts an incident elliptical polarization of Axial Ratio (AR) 5 and tilted φ=20°, into a purely linear vertical polarization of XP<-30 dB with zero reflections and insertion loss better than 0.1 dB, thus improving the electrical performance reported in the literature for these devices. This design has been also manufactured and tested, and the measurements are in good agreement with simulations.

## II. BI-MODE CIRCUIT MODELLING

### A. Theoretical Considerations

Let us consider a square period single layer Planar Periodic Structure ("PPS" or "metasurface"), which is placed in the vacuum at z=0 and is made up a zero-thickness metallizations of arbitrary geometry (Fig. 1a). In these conditions, a complete set of orthogonal modes can be defined for each angle of incidence at both sides of the discontinuity, which are commonly known as "Floquet Modes".

If an incident field impinges normally on the PPS, there are two degenerated modes (zero cut-off frequency) at both sides of the discontinuity: Vertical (V) and Horizontal (H) (Fig. 1b), which can describe an arbitrary incident polarization at normal incidence. High order modes (or harmonics) are also present at both sides of the discontinuity; however, they will be considered evanescent within the band of interest.

Because the metallizations do not show any type of symmetry, an incident V-field (or H-field) will produce vertical and horizontal components at both transmission and reflection, so that the equivalent circuit of the PSS must be modelled as a four-port network whose scattering matrix (using the ports definitions of Fig. 1c) is:

$$S = \begin{bmatrix} S_{11} & S_{12} & S_{13} & S_{14} \\ S_{12} & S_{22} & S_{14} & S_{24} \\ S_{13} & S_{14} & S_{11} & S_{12} \\ S_{14} & S_{24} & S_{12} & S_{22} \end{bmatrix} \quad (1)$$

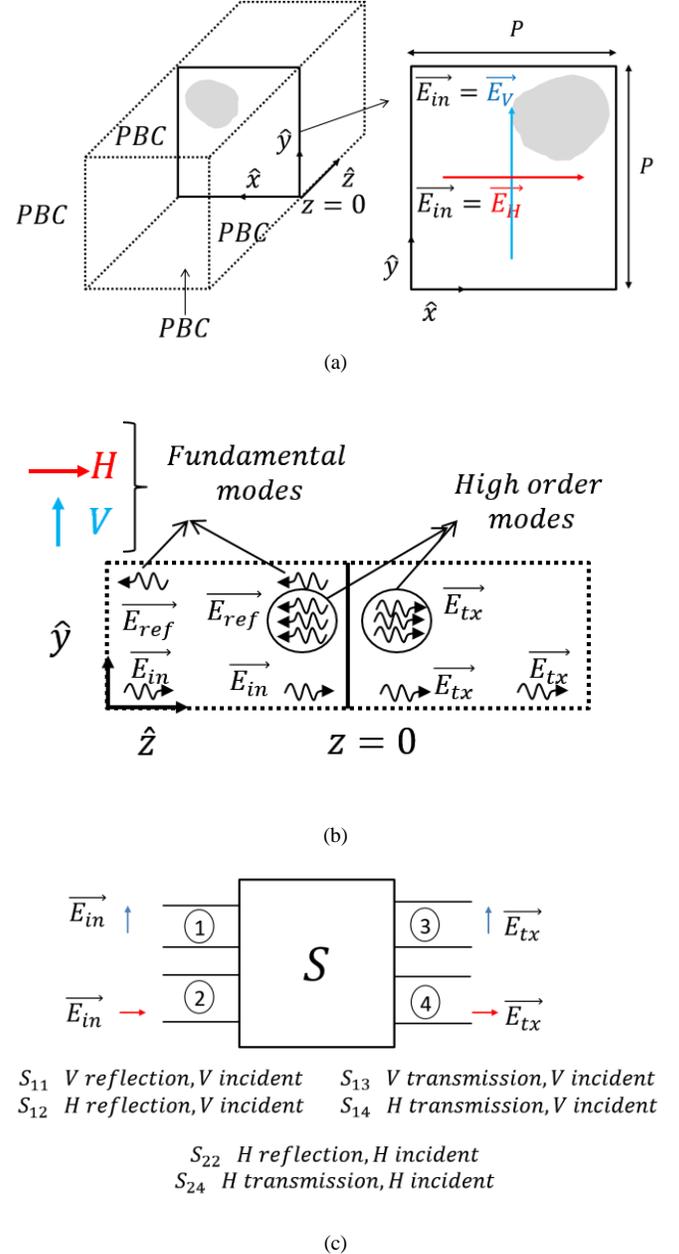

Fig. 1. (a) Architecture of a PPS cell with arbitrary geometry and Periodic Boundary Conditions (PBC), (b) representation of the field components and modes at both sides of the PPS discontinuity, (c) S-parameters and port definitions of the PPS

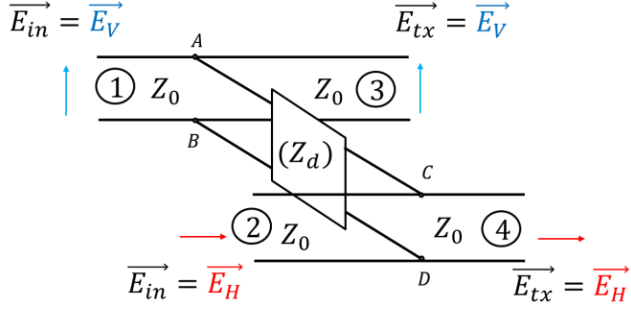

Fig. 2. Bi-mode Foster's Equivalent circuit of an arbitrary PPS for normal incidence

In (1) the input-output symmetry of the structure has been used to reduce the number of independent elements of the matrix.

Since the discontinuity has zero-thickness, the field continuity must be produced mode by mode, so that the following relationships can be obtained:

$$S_{13} = 1 + S_{11} \quad (2)$$
$$S_{14} = S_{12} \quad (3)$$
$$S_{24} = 1 + S_{22} \quad (4)$$

Therefore, it is clear that the scattering matrix of an arbitrary PPS in these conditions can be described by three independent parameters:

$$S = \begin{bmatrix} S_{11} & S_{12} & 1+S_{11} & S_{12} \\ S_{12} & S_{22} & S_{12} & 1+S_{22} \\ 1+S_{11} & S_{12} & S_{11} & S_{12} \\ S_{12} & 1+S_{22} & S_{12} & S_{22} \end{bmatrix} \quad (5)$$

### B. Equivalent Circuit and Canonical Topologies

The distribution of elements in (5) allows the architecture within the box in Fig. 1c to be known, which is the same circuit topology as that shown in Fig. 2. It comprises two transmission lines of the same characteristic impedance (vacuum impedance, $\eta_0=120\pi$), which are interconnected by a purely reactive (the conductor losses are considered negligible) parallel two port network whose impedance matrix $\mathbf{Z_d}$ is:

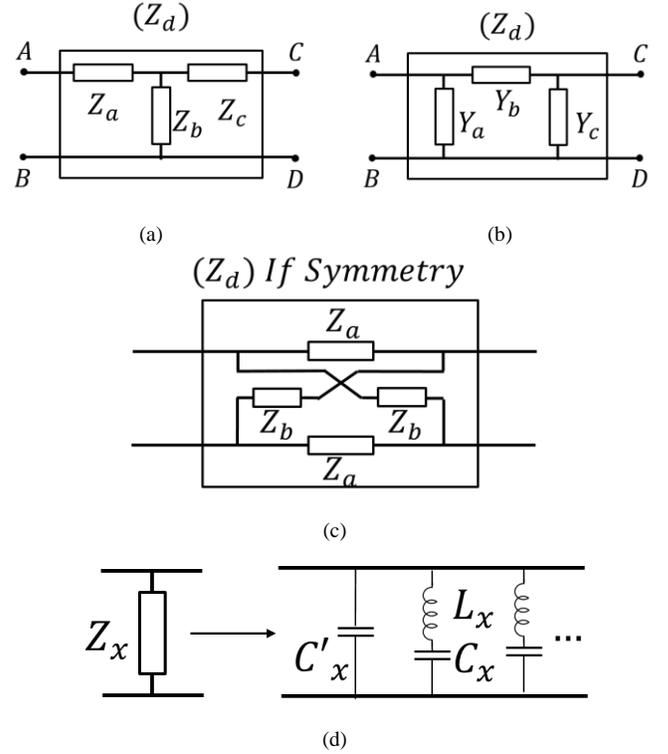

Fig. 3. Canonical topologies for the two port interconnection network, Zd, of the circuit of Fig. 2, (a) T- equivalent, (b) Π- equivalent, (c) Lattice equivalent for the symmetric case and (d) Foster's second form to describe the elements (impedances or admittances) of Zd

$$Z_d = \begin{bmatrix} Z_{d11} & Z_{d12} \\ Z_{d21} & Z_{d22} \end{bmatrix} \quad (6)$$

To obtain the circuital elements that must be included in $\mathbf{Z_d}$, the following classic relationships between $\mathbf{Z}$ and $\mathbf{S}$ matrices are used:

$$Z = \eta_0 [I - S]^{-1} \cdot [I + S] \quad (7)$$

$$S = [Z - \eta_0 I]^{-1} \cdot [Z + \eta_0 I] \quad (8)$$

These equations yield to the next matrix distribution for the impedance of the 4 port network:

$$Z = \begin{bmatrix} Z_{11} & Z_{12} & Z_{11} & Z_{12} \\ Z_{12} & Z_{22} & Z_{12} & Z_{22} \\ Z_{11} & Z_{12} & Z_{11} & Z_{12} \\ Z_{12} & Z_{22} & Z_{12} & Z_{22} \end{bmatrix} \quad (9)$$

where

$$Z_{11} = -\frac{S_{11}S_{22} - S_{12}^2 + S_{22}}{2(S_{11}S_{22} - S_{12}^2)} \quad (10)$$

$$Z_{12} = -\frac{-S_{12}}{2(S_{11}S_{22} - S_{12}^2)} \quad (11)$$

$$Z_{22} = -\frac{S_{11}S_{22} - S_{12}^2 + S_{11}}{2(S_{11}S_{22} - S_{12}^2)} \quad (12)$$

It is interesting to point out that **Z** is ill-conditioned (there is no admittance matrix, Y), as expected from all-parallel structures.

From (9), it is possible to calculate the parameters of (6) by only loading the corresponding ports with an open circuit. Thus, it is obtained that:

$$\mathbf{Z}_d = \begin{bmatrix} Z_{d11} & Z_{d12} \\ Z_{d21} & Z_{d22} \end{bmatrix} = \begin{bmatrix} Z_{11} & Z_{12} \\ Z_{12} & Z_{22} \end{bmatrix} \quad (13)$$

where $\mathbf{Z}_d$ is now well-conditioned.

From (13) and the classic circuit theory, it can be deduced that $\mathbf{Z}_d$ is odd-real positive, so that the three independent elements can be represented by a Foster form (Fig. 3d) as:

$$Z_{ij} = jX_{ij}(\omega) = j\left(-\frac{k_0}{\omega} + 2\frac{k_1}{\omega_1^2 - \omega^2} + \ldots + k_\infty \omega\right) \quad (14)$$

Thus, it can be concluded that a combination of frequency-independent lumped elements allows the full electrical behavior of an arbitrary PPS to be described, provided that the operating frequency is lower than the cutoff of the first high order mode. In the most general case, the PPS does not exhibit any type of symmetry, and the cutoff frequency of the first evanescent mode ($TE_{10}$), and therefore the bandwidth of validity of the circuit (dual-mode bandwidth), is given by:

$$f_{cutoff} = \frac{c_0}{P} \quad (15)$$

where P the period of the unit cell.

The number of terms to be used in (14) depends on the geometry of the PPS and the accuracy to be achieved within the bandwidth of validity of the circuit. It must be highlighted that the Foster's expansion implies the minimum number of circuit elements to describe the reactance, and therefore the simplest circuit form.

Now, any of the classic circuit topologies can be introduced for the two port network (Fig. 3). For a T equivalent form (Fig. 3a), the impedance matrix $\mathbf{Z_d}$ is:

$$\mathbf{Z}_d = \begin{bmatrix} Z_{11} & Z_{12} \\ Z_{12} & Z_{22} \end{bmatrix} = \begin{bmatrix} Z_a + Z_b & Z_b \\ Z_b & Z_b + Z_c \end{bmatrix} \quad (16)$$

and consequently:

$$\begin{aligned} Z_a &= Z_{11} - Z_{12} \\ Z_b &= Z_{12} \\ Z_c &= Z_{22} - Z_{12} \end{aligned} \quad (17)$$

Similarly, the Π equivalent circuit (Fig. 3b) leads to:

$$\mathbf{Z}_d = \frac{1}{Y_aY_b + Y_aY_c + Y_bY_c}\begin{bmatrix} Y_a + Y_b & Y_b \\ Y_b & Y_c + Y_b \end{bmatrix} \quad (18)$$

and

$$\begin{aligned} Y_a &= \frac{Z_{11} - Z_{12}}{Z_{11}Z_{22} - Z_{12}^2} \\ Y_b &= \frac{Z_{12}}{Z_{11}Z_{22} - Z_{12}^2} \\ Y_c &= \frac{Z_{22} - Z_{12}}{Z_{11}Z_{22} - Z_{12}^2} \end{aligned} \quad (19)$$

It should be highlighted that the parameters $Z_{11}$ and $Z_{22}$ are positive-definite (their residues are real-positive) so that the slope of the reactance curves is also positive. However, the reactance of $Z_{12}$ can be positive or negative. Consequently, the elements of the equivalent T or Π forms could show negative values of inductances and capacitances, which do not invalidate the use of the equivalent circuit and neither infringes the fundamental postulates that support the circuit theory [36].

In the particular case that the PPS geometry presents a symmetry with respect to one of the two diagonal axes ($S_{11}=S_{22}$), the lattice topology (which always provides positive lumped elements) can be introduced (Fig. 3c). The parameters are:

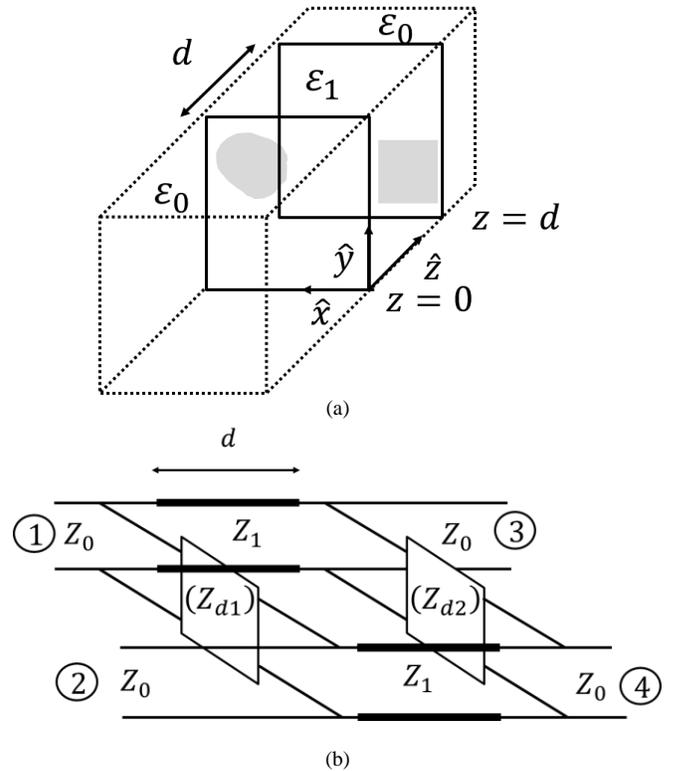

Fig. 4. (a) Architecture of two stacked PPS cells of arbitrary geometry that are separated by a dielectric spacer of permittivity ($\varepsilon_1$), (b) Bi-mode Equivalent circuit for normal incidence

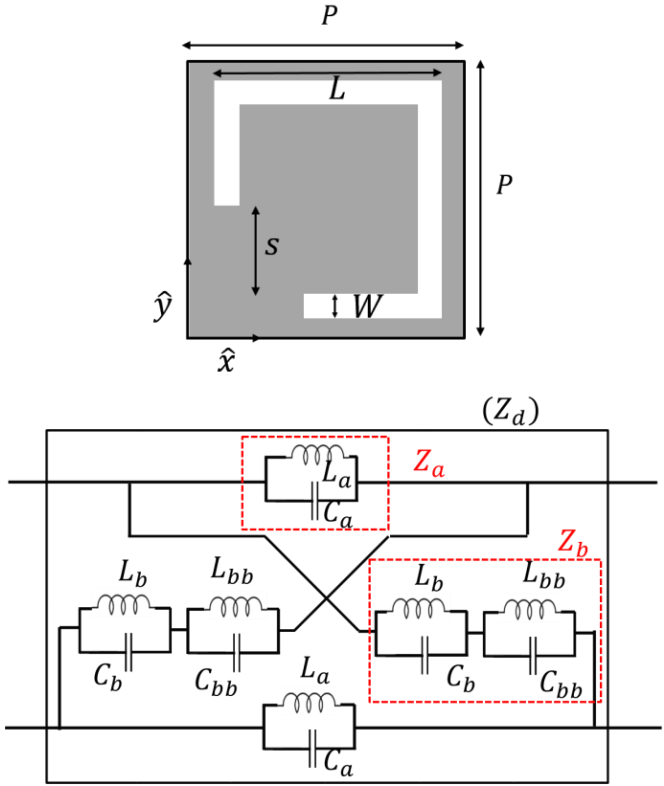

Fig. 5. Top view of a PPS based on a defected slotted ring, and its lattice equivalent form for the two port interconnection network of Fig. 2

$$Z_d = \begin{bmatrix} Z_a + Z_b & Z_b - Z_a \\ Z_b - Z_a & Z_a + Z_b \end{bmatrix} \quad (20)$$

and

$$Z_a = \frac{Z_{11} - Z_{12}}{2}$$
$$Z_b = \frac{Z_{11} + Z_{12}}{2} \quad (21)$$

Note that the circuit presented in [12], is a particular case that can be described with the lattice form.

It is very important to point out that the equivalent circuit presented here is valid for an arbitrary geometry of the PPS and period. A change of the geometry (dimensions or shape) just produces the variation of: the number of poles, the values of their frequency positions and the residues.

### C. Generalizations of the Circuit

The circuit presented in Fig. 2 can be generalized in the case that arbitrary lossless dielectrics are placed on both sides of the PPS discontinuity (Fig. 4a). The circuit elements are different than those obtained when considering the vacuum, and must be calculated by simulating the scattering parameters of the PPS in the presence of the dielectrics. The final equivalent circuit results in two pairs of sections of transmission lines placed at the corresponding ports, which must have the appropriate impedances and physical lengths (thickness), as shown in Fig. 4b.

Other generalization is allowed for rectangular periods simply by multiplying the impedances of the ports by the factors ($\sqrt{Py/Px}$) and ($\sqrt{Px/Py}$).

The electrical effects of a non-zero metal thickness can be also accounted in the circuit if two air transmission line sections are included, one on each side of the connecting two-port network, provided that the thickness of the metals ($t_m$) is a little fraction of the wavelength ($t_m<\lambda/20$). Otherwise, the metals would form a new periodic transmission line of length $t_m$, whose particular set of modes differ from those of the Floquet [39]. Consequently, (2)-(4) are not fulfilled, and other relations must be found to describe an equivalent circuit. In [13], equivalent

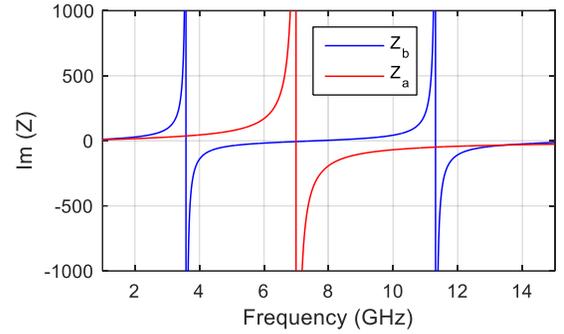

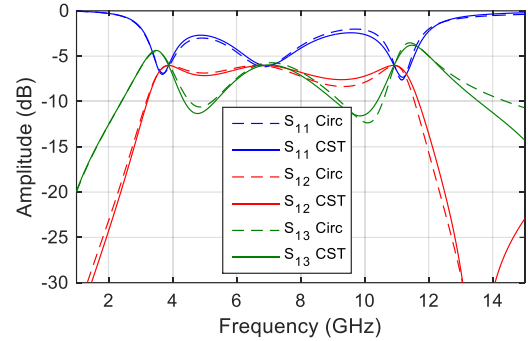

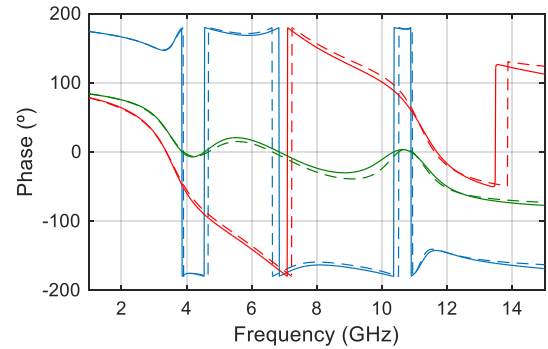

Fig. 6. Simulation of the PPS cell shown in Fig. 5 (P=15 mm, W=0.5 mm, L=13 mm and S=3 mm) using both the circuit and CST, (a) Reactances of the lattice form, (b)-(c) Amplitude and Phase of the parameters: $S_{11}$, $S_{12}$ and $S_{13}=1+S_{11}$,

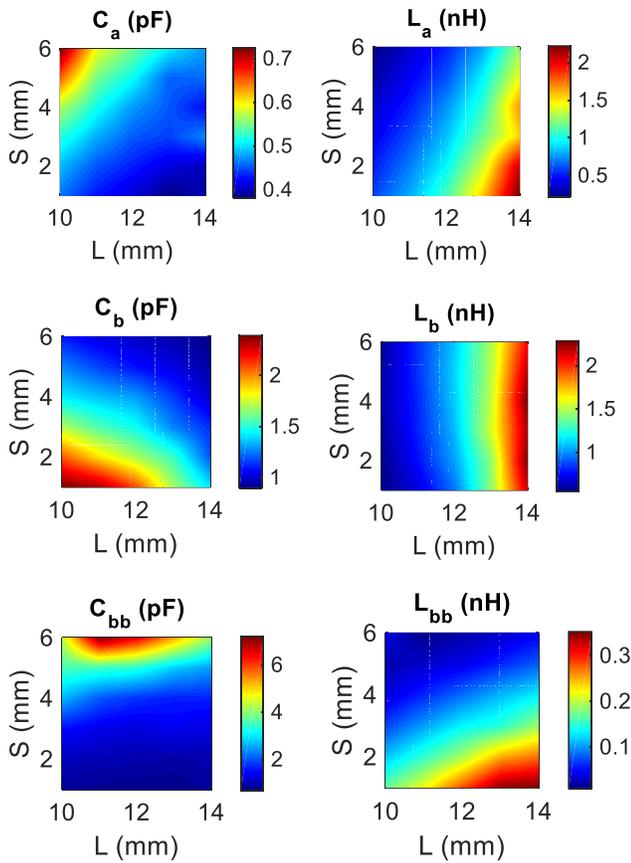

Fig. 7. Values of the lumped elements of the PPS of Fig.5 as a function of the dimensions, L and S. P=15 mm and W=0.5 mm

circuits of these types of periodic architectures (commonly known as 3D) are presented.

Once a PPS is fully characterized using the circuit presented in Fig. 2, a multi-layer planar structure comprising different PPSs and dielectric spacers can be easily analyzed by multiplying the classic four-port transmission (T) or (ABCD) matrices of each component (PPSs and dielectrics) [40].

### D. Advantages and Limitations

The equivalent circuit proposed here has certain advantages when compared to other circuits reported in the literature to model planar 2-D periodic structures [22]-[35]: (a) the topology is invariant to the PPS geometry and makes it possible to use classic circuit theory to design a multitude of devices, (b) the process to obtain the circuit can be standardized, (c) the Foster form implies the simplest circuit in terms of elements [36], and (d) the circuit can be completely described with lumped elements that are independent of the frequency. This latter allows for important additional benefits such as saving storage space related with the frequency and the possibility to use the scaling factor (k). Thus, if all the dimensions of a cell (scatter and period) are multiplied by k, the values of the new capacitances and inductances are calculated just by multiplying and dividing by k, respectively.

When compared to other mathematical tools that can be used to model the electrical behavior of a PPS (i.e. Artificial Neural Networks, ANN [41] or Supporting Vector Machines [42]), the aforementioned advantages also result in improving the accuracy and the ratio between the time to obtain the circuit versus the training time of the mathematical tools, as a sophisticated spectral behavior of the PPS can be the limiting factor to make them unproductive. Moreover, these tools have no information about the physical behavior of the structure, so a multi-layer structure must be simulated and trained even when the PPSs that compose it had been previously and individually characterized. It is important to note that although these tools could be considered as a black box providing the S parameters of the PPSs, certain circuit aspects (and therefore physical ones) must be considered to cascade them and describe a multi-layer structure: the existence of modes at the ports and its physical propagation. However, even when compared with these mixed strategies, the proposed circuit presents the advantage to extract the frequency.

It should be also highlighted that the proposed circuit is especially appropriated to perform design procedures for devices that allow a standardized synthesis among those studied in the classical circuit theory (filters, LP-CP converters, which can use the coupler theory [12], transmitarray cells, etc). Note that a standardized synthesis process provides relevant information about the electrical limits of the devices to be designed in terms of canonical circuit topologies and distributed elements (i.e the minimum number of layers required to reach certain electrical specifications). Moreover, these processes provide the values of the elements that must be achieved at each layer in terms of Foster's elements. To develop other type of devices, the proposed circuit can also aid to perform novel design strategies that require of an efficient optimization. In these cases, the equivalent circuit combined with an EM

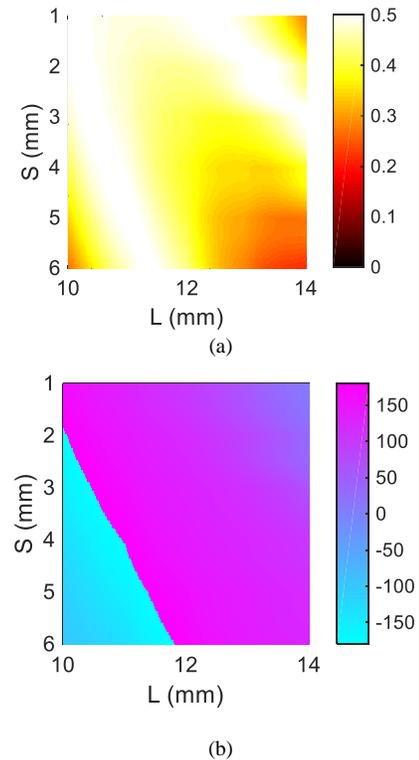

Fig. 8. $S_{12}$ of the cell of Fig. 5 as a function of the parameters, L and S. P=15 mm and W=0.5 mm, at 10 GHz.

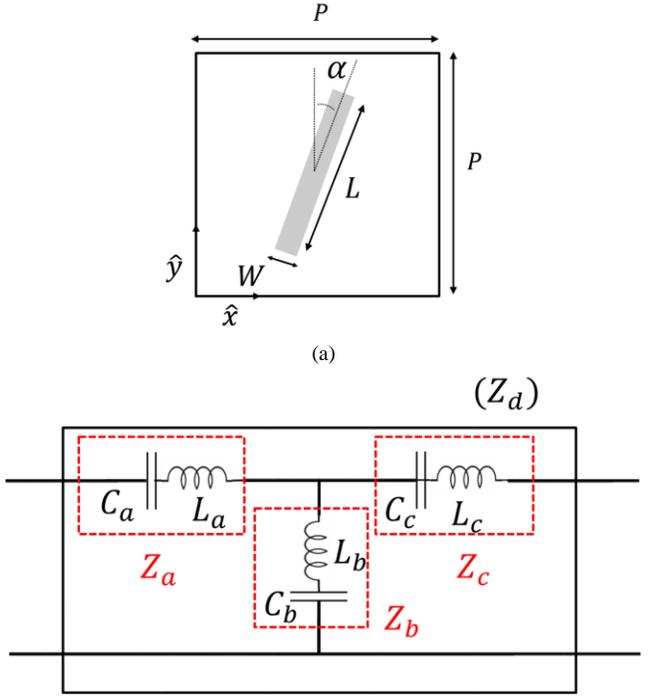

Fig. 9. (a) Top view of a PPS made up a rotated dipole, and (b) its T equivalent form for the two port interconnection network of Fig. 2.

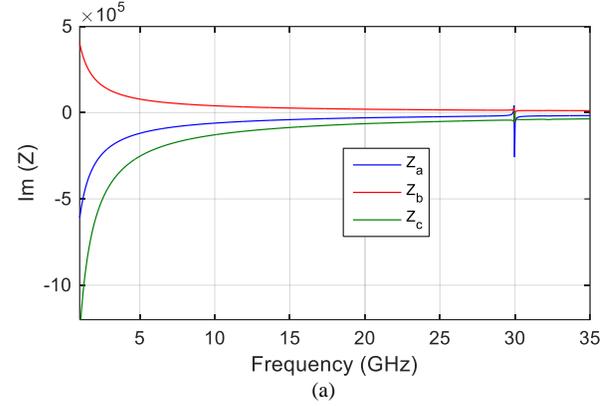

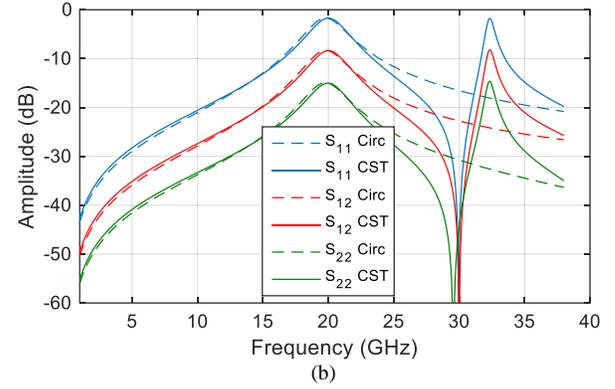

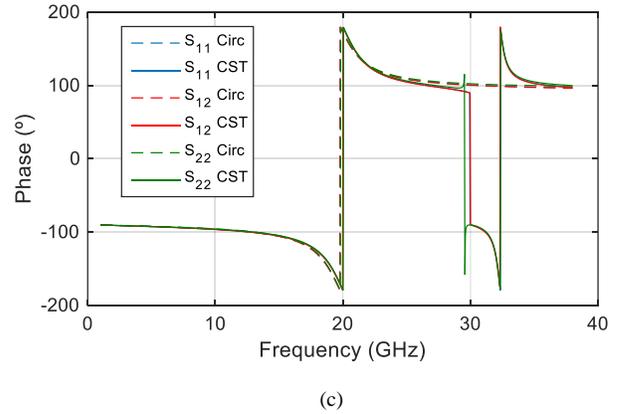

Fig. 10. Simulation of the PPS cell shown in Fig. 9 (P=10 mm, W=0.5 mm, L=7 mm and α=25º) calculated using both the circuit and CST, (a) Reactances of the T equivalent, (b)-(c) Amplitude and Phase of the parameters: $S_{11}$, $S_{12}$ and $S_{22}$

simulator can play an important role, especially to optimize those devices made up of a large number of layers or cells.

As regards the drawbacks, two main limitations are clearly identified. On the one hand, the proposed circuit is only valid within the bi-mode bandwidth of the PPS. Although most of the devices work in this band, it can be possible that the device includes thicknesses and/or permittivities for which the evanescent modes may not be enough attenuated (at least 10 dB) between discontinuities, so the accuracy of the circuit can be drastically deteriorated. Therefore, other circuits should be considered in these cases (i.e. [27], [34]]. It must be pointed out that if dielectrics spacers are used, the high order modes that are evanescent in the vacuum could be propagating within the dielectrics. Therefore, the bandwidth given by (15) must be strictly divided by $\sqrt{\varepsilon_{rh}}$ ($\varepsilon_{rh}$ is the higher permittivity used in the structure). However, a multilayer structure that uses the dielectrics to partially fill the volume between discontinuities or is comprised of scatters that do excite the high order modes with enough amplitude, could overcome this new limit. Therefore, the functional bandwidth of the proposed circuit will depend on several factors: the permittivities and thicknesses of the dielectrics, the geometries of the scatters, and the accuracy required for the design.

On the other hand, in spite of the generality of the circuit (no symmetries are required for the PPS geometry), it has been only studied and validated for normal incidence. The normal incidence can be used to perform a large number of devices or even be a good starting point to design devices at oblique incidence.

It should be noted that if the proposed circuit is able to model the absence of physical symmetries, it could be possible that the absence of symmetries in the excitation (oblique incidence) were also appropriately modelled. Some recent results we have obtained make us very optimistic about the capability of the proposed circuit to model an arbitrary oblique incidence, provided that the appropriate basis of modes (impedances and propagation constants) are selected at each port. This analysis will be addressed in further works.

### III. NUMERICAL VALIDATION

In this section, the equivalent circuit presented above is numerically evaluated by analyzing two different PPS architectures that have been used in the literature to develop mm-wave devices.

## A. Example 1: Defected Slotted Ring

The first geometry is that shown in Fig. 5, which consists of a defected slotted ring on a metallic sheet. This element was introduced in [43] to develop a single–layer transmit-reflectarray antenna, which takes advantage of the properties of the cross-polar (XP) field to obtain a full phase-range, thus avoiding the limitations of the CP in terms of number of layers [20]. The element presents four independent physical parameters (L, W, S and P), although the width of the slot and the period will be fixed as in [43]: W=0.5mm and P=15 mm. Since the cell presents one symmetry with respect to a diagonal axis, the lattice topology will be selected to describe the equivalent circuit, although the other general topologies (T or Π) could be also used.

Firstly, the parameters of the equivalent circuit have been obtained for the dimensions (L=13 mm and S=3 mm), to evaluate the accuracy of the equivalent circuit in terms of bandwidth. Thus, the S parameters were calculated from 1 to 15 GHz using the EM simulator CST [44], from which the elements of the lattice, Za and Zb, were obtained.

Fig. 6a shows the spectral behavior of the reactances of Za and Zb. It can be seen that Za has a single pole, so a single shunt resonator, whose parameters are (Ca, La), is enough to describe its electrical behavior with a reasonable accuracy. In contrast, Zb exhibits two poles, and therefore two series-connected parallel resonators are required: (Cb, Lb) and (Cbb, Lbb). Fig. 5 shows the circuit for the two port interconnection network.

To calculate the values of the lumped elements, any curve fitting technique can be used. The poles of Zb are quite separated, which helps to find the pairs (Cb=1,259 pF, Lb=1,47 nH) and (Cbb=1,355 pF, Lbb=0,147 nH). The calculus of the parameters (Ca=0,442 pF, La=1,172 nH) does not suppose any significant problem.

Fig. 6b and Fig. 6c show a comparison between the S parameters of the cell in the band from 1 to 15 GHz, which are calculated using both the equivalent circuit and the EM-simulator. As can be noted, the circuit predicts the electrical behavior of the PPS with excellent accuracy in the whole band. Obviously, additional resonators can be included to obtain even a better accuracy (especially in the higher frequencies of the band of interest) or if the operating band were required to be wider.

Since a dimensional variation of the PPS does not change its geometry drastically, the same spectral behavior shown in Fig. 6a can be assumed for every set of dimensions. Fig. 7 details the values of the lumped elements as a function of the geometrical variables L and S, where 5x5=25 samples were obtained from electromagnetic simulations (they are corresponded with the values: L=10, 11, 12, 13 and 14 mm and S=1, 2 3, 4 and 5 mm). The values of the lumped elements for other intermediate dimensions were interpolated. Once the elements of the circuit are calculated, it is easy to obtain the S parameters of the PPS

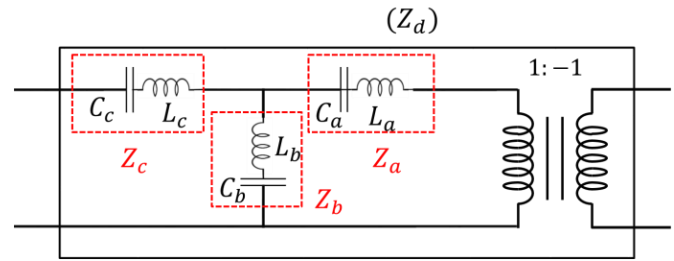

Fig. 12. Equivalent circuit of the two port interconnection network of the PPS in the case that the structure of Fig. 9 is rotated 180º with respect to the x-axis

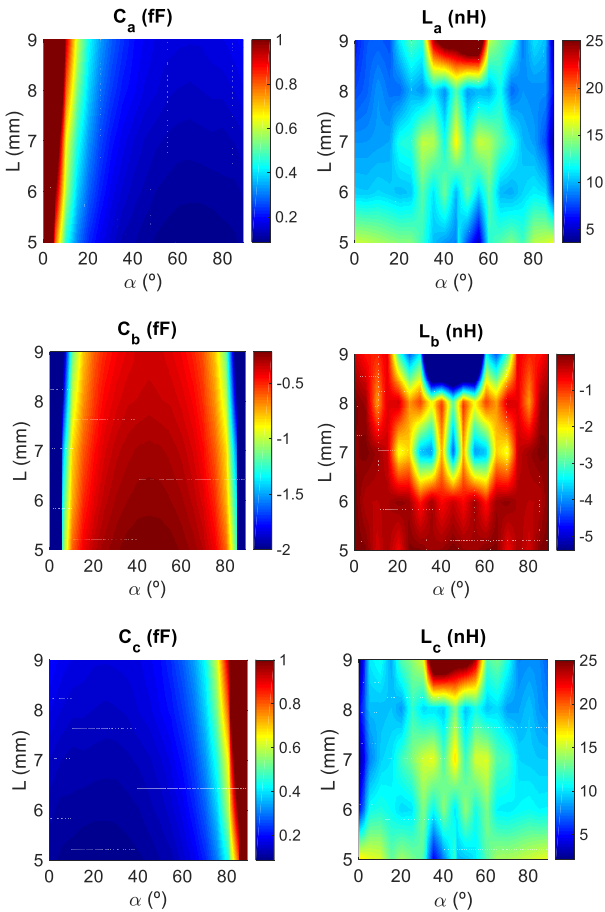

Fig. 11. Values of the lumped elements of the PPS of Fig. 9 as a function of dimensions, L and α. P=10 mm and W=0.5 mm

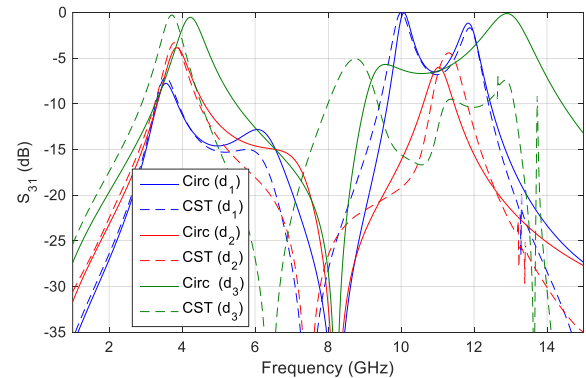

Fig. 13. Simulated $S_{31}$ for the two-layer design presented in section III.C (Example 3), where several separations are accounted. The circuit and EM simulations (CST) are compared

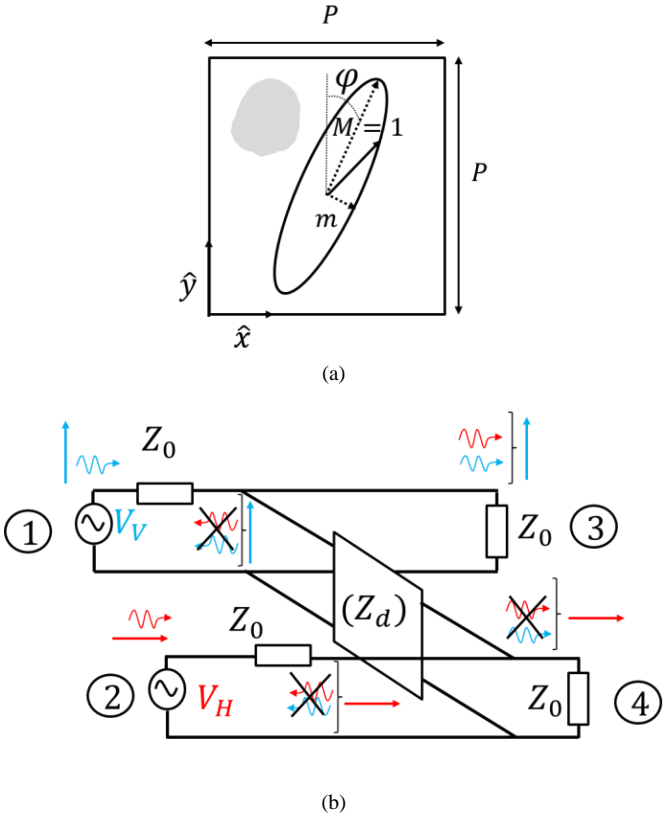

Fig. 14. (a) Architecture of a PPS cell illuminated by an arbitrary elliptically polarized field (the maximum field is assumed unitary, M=1, without loss of generality), and (b) equivalent circuit, which shows the conditions to convert and transmit the field with a purely vertical polarization

at every set of dimensions and frequency using (8). For instance, the parameter S12 was calculated at 10 GHz for every dimensional value. The results given by the circuit in Fig. 8 can be compared with those provided in [43].

### B. Example 2: Rotated Dipole

The second geometry analyzed is shown in Fig. 9, which consists on a dipole rotated an angle α with respect to the y axis. This element has been used to develop a wide range of devices, especially to improve the XP of planar antennas [45]-[46].

The element shows four independent physical parameters (L, W, α and P), although the width and the period will be fixed again to simplify the analysis: W=0.5 mm and P=10 mm. Because the cell does not exhibit any type of symmetry, the T-equivalent topology was selected for the two-port interconnection network. Fig. 10a shows the impedances Za, Zb and Zc from 1 to 35 GHz in the case that α=25° and L=7 mm, which show a highly capacitive behavior. Therefore, the reactances could be modelled just using a single series resonator, thus obtaining the elements represented in Fig. 9. It can be appreciated that a pole appears exactly at the cutoff frequency of the geometry ($f_{cutoff}$=30 GHz), which would imply the use of an additional resonator in the equivalent circuit to accurately predict the electrical behavior at frequencies close to 30 GHz.

The values of the lumped elements were calculated, resulting in: (Ca=0,2724 fF, La=12,203 nH) (Cb=-0,409 fF, Lb=-1,893 nH) and (Cc=0,123 fF, Lc=13,072 nH). Fig. 10b and Fig. 10c show a comparison between the S parameters using both the equivalent circuit and the EM-simulator. As can be noted, the accuracy of the circuit is excellent in the band from 1 to 23 GHz. However, it can be appreciated that the circuit cannot predict the spectral behavior from 23 GHz to 30 GHz with enough accuracy, as additional Forster's elements are required (one more element in this case).

Fig. 11 shows the values of the lumped elements as a function of the geometrical variables: L and α. The electromagnetic simulations and therefore the lumped elements were calculated for 50 samples: L=5, 6, 7, 8 and 9 mm and α=0, 5, 10…45° (steps of 5°). It can be appreciated that Cb and Lb tend to ∞ and 0 (Zb=0), as the angle is close to 0° and 90° respectively, which would result in an independent circuit for each polarization, as expected.

Note that that the values of lumped elements in the range (α=46° to 90°) can be obtained from those calculated from α=0° to 45°, since (Za, Zc) interchange to each other with respect to the diagonal the cell. Similarly, the values from α=91° to 179° (α=-89° to -1°) can be also easily deduced from those previously calculated just by considering the mirror structure of Fig. 9 with respect to the x-axis. This case can be characterized in the circuit if Za and Zc are interchanged and if an additional transformer that accounts the change of sign of the field is included, as shown in Fig. 12. Once the elements are calculated

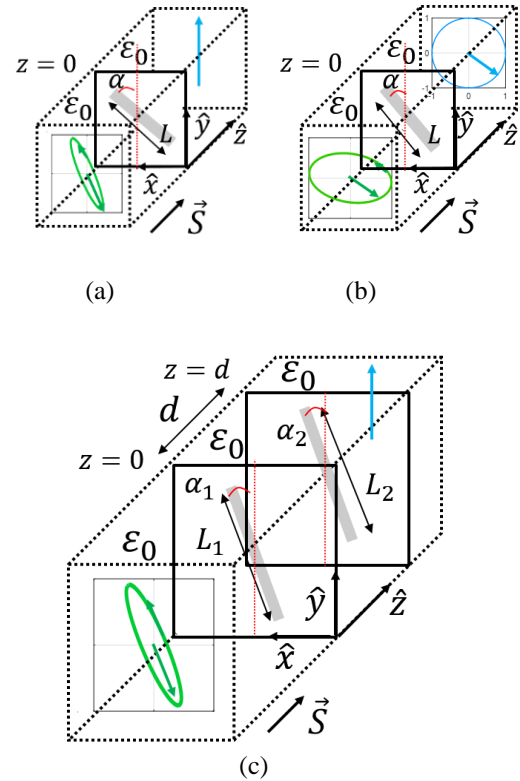

Fig. 15. Architecture of several polarization controllers based on cells made up of rotated dipoles: Single Layer EP-LP converter, (b) Single Layer EP-CP converter, and (c) Two Layer EP-LP converter

in the presence of the transformer, the new lumped elements of the T-equivalent can be easily obtained.

It is important to note that the calculus of the inductances (La, Lb, and Lc) is very sensitive in this case because of the capacitive behavior of the rotated dipole. However, the inductances cannot be neglected from the equivalent circuit, as they are important to appropriately describe the frequency behavior of this PPS. Therefore, particular attention should be paid to the meshing of the EM simulations or the accuracy of the fitting strategy used to calculate the elements.

### C. Example 3: Two Stacked Defected Slotted Ring

As an example to evaluate a multi-layer cell, the structure comprising the PPS shown in Fig. 5 and its mirror equivalent cell is considered at 10 GHz. It is easy to see that the lattice equivalent circuit of the mirror cell can obtained from that of the original one by interchanging Za and Zb. To perform the analysis, both elements are separated by an ideal vacuum spacer of three different thicknesses: $d_1=\lambda_0/2$, $d_2=\lambda_0/7$ and $d_3=\lambda_0/10$. In the first case ($d_1=\lambda_0/2$), the resulting circuit will be composed of two lines, which are interconnected by a two-port network that results from the parallel connection of a ladder network and other one with its branches interchanged. The S parameters of this particular circuit architecture ($S_T$) are known at $f_0$:

$$S_{11T} = -2 \frac{S_{11}^2 - S_{12}^2 + S_{11}}{3S_{11}^2 - 3S_{12}^2 + 4S_{11} + 1}$$

$$S_{12T} = S_{14} = 0 \qquad (22)$$

$$S_{13T} = 1 + S_{11T}$$

which are written as a function of the S parameters of the original cell. It can be noted that no cross polarization is produced at $f_0$ for an arbitrary set of dimensions. Therefore, the parameters of Fig. 7 can be used to calculate $S_T$ and to identify the dimensions that cancel the reflection $S_{11T}$, thus performing for example a passband filter. This condition is fulfilled if Za=-Zb, so that the dimensions (L=13.4 mm and S=3 mm) are obtained and will be used to carry out the multi-layer analysis. Fig. 13 shows the parameters $S_T$ for this simple example calculated using the equivalent circuit and CST. The results for the other thicknesses are also represented. As can be seen, it can be noted that the equivalent circuit is able to predict the electrical behavior of a multi-layer structure with tolerable accuracy even when thinner spacers are used. The presence of evanescent modes is appreciable when the thicknesses are lower than $\lambda/7$, although in general it will depend on the PPS geometry and the dielectrics used, as mentioned above.

## IV. DESIGN CASES AND EXPERIMENTAL VALIDATION

This section is focused in the study and development of devices capable of converting an arbitrary elliptical polarization into a purely vertical linear or circular polarization (EP-LP or EP-CP converters), thus combining the rotation and the cancelation of field components at the same time. These devices are selected to demonstrate the accuracy of the circuit in order to efficiently perform complex polarization controller devices.

### A. Design 1: Single Layer Transmissive Elliptical to Linear Polarization (EP-LP) Converter

The first design consists on a transmissive (EP-LP) converter that operates at 20 GHz, which is made up of a single layer PPS. Fig. 14a represents the general parameters of an arbitrary elliptically polarized field that impinges normally on the PPS with a maximum unitary field (M=1). The incident field is:

$$E_{iV} = (\cos\phi \pm jm\sin\phi) \qquad (23)$$

$$E_{iH} = (\sin\phi \mp jm\cos\phi) \qquad (24)$$

where ($\pm$) indicates a Left Hand (LH, +) or Right Hand (RH, -) polarization.

In this particular case, a LH elliptically polarized incident field tilted $\varphi=20º$ and whose axial ratio is 5 (M=1 and m=0.2), is considered at 20 GHz (see Fig. 15a). From the circuit point of view, the design of EP-LP converters requires the use of two independent generators set at ports 1 and 2, which represent the contribution of the elliptically polarized field to the vertical and horizontal incident components. Thus, a full design should cancel the reflections at ports 1-2 and the transmission to port 4. The reflection level at port 1 (R1) is a sum of the own vertical reflection of the PPS ($S_{11}$) and the cross reflection between ports 1 to 2 ($S_{12}$), as depicted in Fig. 14b. Thus, $R_1$ can be written as:

$$R_1 = (\cos\phi \pm jm\sin\phi)S_{11} + (\sin\phi \mp jm\cos\phi)S_{12} \qquad (25)$$

Similarly, the reflection fields at port 2 ($R_2$) and the

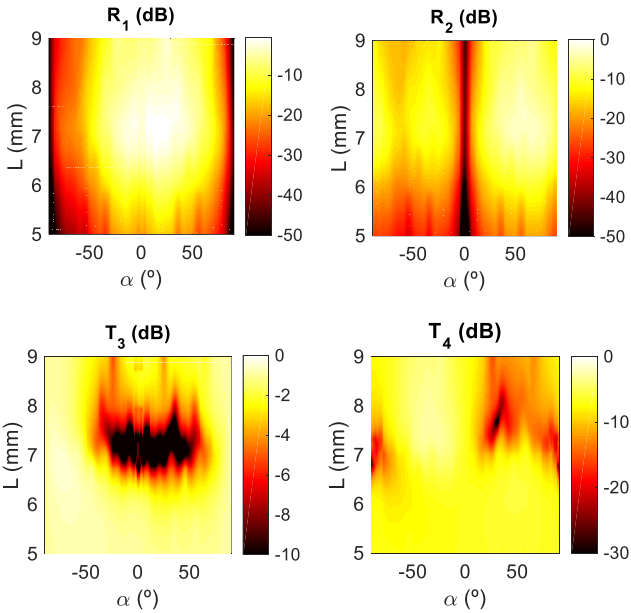

Fig. 16. Reflections and transmissions (20 GHz) at each port calculated using the circuit for a single layer PPS based on a rotated dipole cell (Fig. 15a). The parameters are plotted as a function of L and α, in the case that a LH elliptically polarized incident field (φ=25º, M=1, m=0.2) impinges at normal incidence.

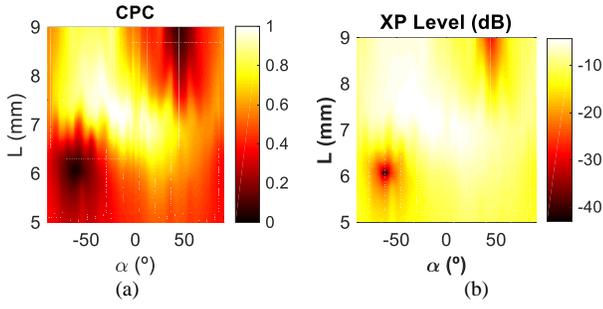

Fig. 17. (a) Circular polarization condition and (b) XP level at 20 GHz as a function of both the dipole length L and the rotation angle α (Fig. 15b), which is used to design the EP-CP converter of section IV.B. LH-EP incident field (φ=80°, M=1, m=0.6).

transmissions to ports 3 ($T_3$) and 4 ($T_4$) are:

$$R_2 = (\cos\phi \pm jm\sin\phi)S_{12} + (\sin\phi \mp jm\cos\phi)S_{22} \quad (26)$$

$$T_3 = (\cos\phi \pm jm\sin\phi)(1+S_{11}) + (\sin\phi \mp jm\cos\phi)S_{12} \quad (27)$$

$$T_4 = (\cos\phi \pm jm\sin\phi)S_{12} + (\sin\phi \mp jm\cos\phi)(1+S_{22}) \quad (28)$$

where the relations (1)-(4) are used to substitute the parameters $S_{13}$, $S_{23}$, $S_{24}$, etc.

It can be deduced from (23)-(28) that $R_1=R_2=T_4=0$ cannot be fulfilled at the same time, as $R_2 = T_4 - \sin\phi \mp jm\cos\phi$. Therefore, a complete design cannot be performed just by using a single PPS layer. However, it is possible to cancel ($T_4$) at the expense of assuming a certain level of reflection.

Because these conditions require of non-symmetrical PPSs, the structure used for the design is that shown in Fig. 9. Fig. 16 represents the amplitude of the reflections and transmissions waves calculated with the equivalent circuit at each port as a function of the physical parameters of the PPS (L and α). The amplitudes $R_1$, $R_2$, $T_3$ and $T_4$ have been normalized by $\sqrt{1+m^2}$, which accounts the incident power. Therefore, $T_3$ evaluates the insertion loss of the device (change of sign), $R_1$ and $R_2$ are the reflection levels and $T_4$ is related to the XP-level.

Two different solutions that cancel the transmission to port 4 ($T_4$) can be appreciated in Fig. 16: (L=7.68 mm, α=29.9°) and (L=6.9 mm, α=90°), which exhibit the levels ($R_1$=-2.26 dB and $R_2$=-6.8) and ($R_1$=-50.1 dB and $R_2$=-6.8 dB), respectively. The insertion loss are 7.52 dB and 1 dB. Thus, it can be noted that a better insertion loss is obtained in the case that the dipole directly eliminates the incident horizontal components of the field at port 2 (α=90°), where $S_{22}\approx-1$ and $S_{11}\approx S_{12}\approx 0$. This latter intuitive case has been widely studied and used in mm-wave and optical frequencies to develop polarizers, although at the expense of assuming a certain reflection level.

### B. Design 2: Single Layer Transmissive (EP-CP) Converter

To perform a transmissive (EP-CP) converter, the transmitted field must fulfill the circular polarization condition:

$$CPC = |E_{tH} \pm jE_{tV}| = |T_4 \pm jT_3| = 0 \quad (29)$$

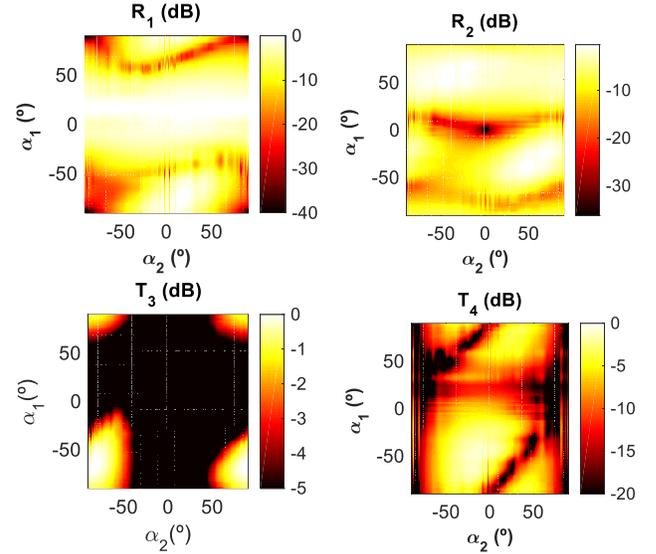

Fig. 18. Reflections and transmissions at each port calculated using the equivalent circuit of the two stacked layer PPSs based on rotated dipoles (Fig. 15c). The parameters are plotted as a function of the rotation angles ($\alpha_1$ and $\alpha_2$), in the case that a LH elliptically polarized incident field (φ=25°, M=1, m=0.2) impinges at normal incidence. The dipole lengths are fixed to ($L_1=L_2=7$ mm).

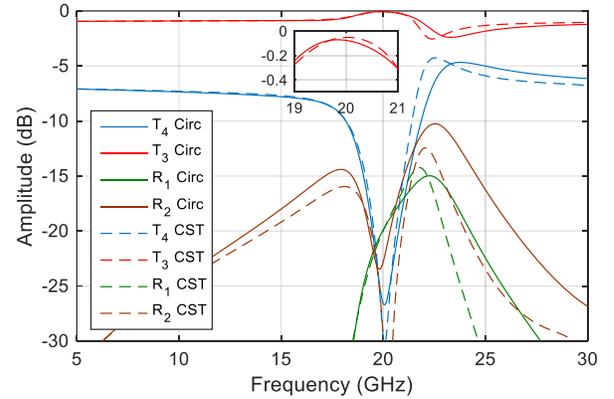

Fig. 19. Spectral behavior of the reflections and transmissions at each port for the device optimized in section IV.C (Fig. 15c), which is calculated using both the circuit and CST.

A LH elliptically polarized incent wave of (M=1, m=0.6) and φ=80° is considered (see Fig. 15b) as a design case. Fig. 17a represents the value of (29) at 20 GHz versus the physical parameters of the PPS (L and α) calculated using the equivalent circuit. It can be seen that this condition is accomplished when L=6.1 mm and α=-61.13° (see Fig. 17a), so that a purely LH circular polarization is transmitted for this set of dimensions. Fig. 17b shows the XP level versus L and α, which is calculated using the conventional formulation and the appropriate normalization. It can be appreciated that the XP level achieved in the design point is -42.84 dB. The reflections at port 1 and 2 have been also calculated, resulting in -16.03 dB and -10.12 dB respectively, and the insertion loss is lower than 0.6 dB (88% of the incident power is transmitted).

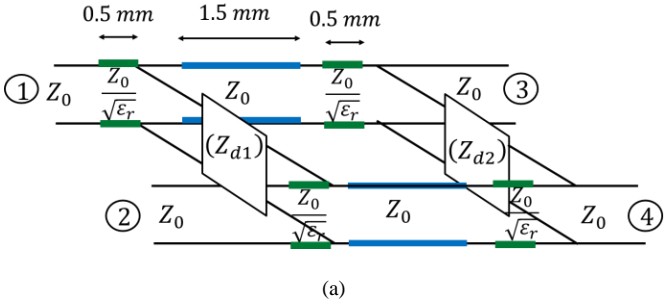

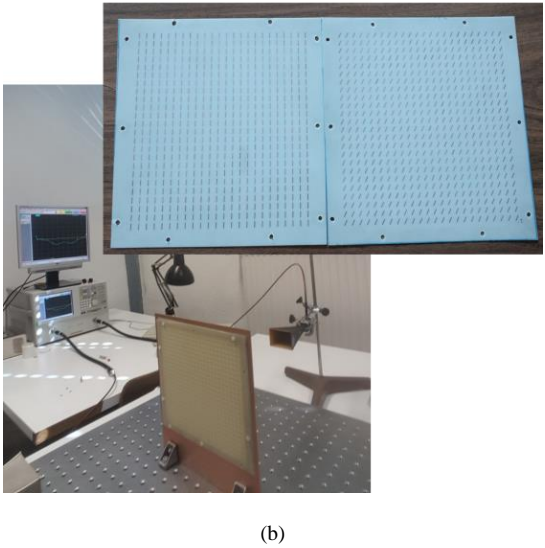

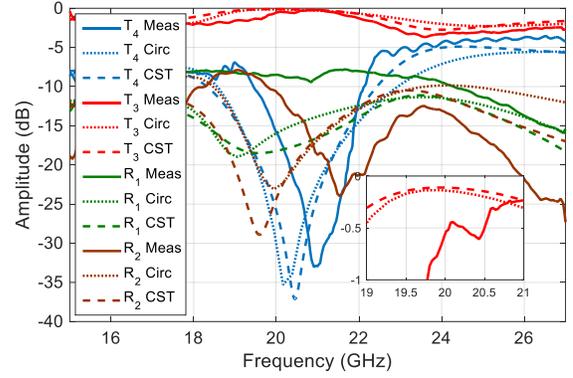

Fig. 21. Spectral behavior of the reflections and transmissions at each port of the manufactured two-layer EP-LP converter. Measurements versus simulations: Electromagnetic (CST), and Circuit in accordance with Fig. 20a

Fig. 20. (a) Equivalent circuit of the structure used to perform the EP-LP converter designed in section IV.D, which includes thin dielectrics, and (b) Manufactured EP-LP converter placed in the system used to measure the S parameters. The $S_{31}$ can be seen in the VNA.

### C. Design 3: Dual Layer Transmissive (EP-LP) Converter

As explained in section IV.A, more than one PPS layer are required to accomplish $R_1=R_2=T_4=0$, as the relations (1)-(4) are not generally imposed in a multi-layer structure. In this section, a two layer EP-LP converter is analyzed and designed. The device is shown in Fig. 15c, which is made up of two different PPSs (that present the geometry of Fig. 9) separated by an ideal vacuum spacer of thickness "d".

In accordance to our best knowledge, the circuit theory does not provide a standardized circuit synthesis that allows to design this device when two generators are used, contrary to what happens with the LP-CP converters [12]. However, we can either continue to use the equivalent circuit and the results obtained in section III to efficiently calculate and optimize the structure.

The design process has five variables: the lengths, rotations and the separation (P=10 mm and W=0.5 mm are fixed), which is complex for a pure electromagnetic optimization. In accordance with the results in Fig. 5, a length of 7 mm will be previously considered for both dipoles to make sure that the resonance is close to 20 GHz. Moreover, the properties of a $\lambda_0/4$ spacer may be a good starting thickness to match the ports even when two generators are used, as the circuit is only made up of shunt elements. Under these conditions, a first iteration gave the results in Fig. 18, where the normalized amplitudes of the reflections and transmissions waves at each port are depicted at 20 GHz as a function of $\alpha_1$ and $\alpha_2$. As can be seen, the first iteration reaches a good point if $\alpha_1$=-65.5º and $\alpha_2$=-86º, for which the insertion loss has been reduced to $T_3$=0.3dB. Then, an optimization routine is directly applied to the circuit to maximize T3 using the starting point calculated above. It leads to the sets of dimensions ($L_1$=6.45 mm and $\alpha_1$=-60.5) and ($L_2$=6.75 mm and $\alpha_2$=-87º), which provides a cross polar level better than XP=-30 dB, and reflections of $R_1$=-19.8 dB and $R_2$=-23.2 dB. The insertion loss is lower than 0.1 dB, thus obtaining a much better result than that obtained using a single layer. Other thicknesses were investigated with the aid of the circuit, which showed that an appropriate thickness should be close to $\lambda_0/4$=10 mm.

Fig. 19 also depicts a comparison between the spectral behavior of the designed EP-LP converter when it is calculated using both the circuit and CST. The circuit simulations have provided an excellent starting point to perform the final design, which will require to use an electromagnetic simulator. A new design involving a different frequency, other materials (with the appropriate calculus of the new lumped elements for them), other input data and/or more layers, can be carried out without additional computational efforts using the proposed equivalent circuit. Thus, the proposed circuit may be a reasonable pre-design tool for these devices, especially to those involving a large number of layers for which an EM simulator could be inefficient, as aforementioned.

### D. Manufacturing and Test

The results and the viability of the proposed equivalent circuit as part of the design procedure of complex polarization controller devices have been demonstrated previously, as the results were compared with those calculated by an electromagnetic simulator (CST). However, an experiment must be carried out to evaluate if the proposed PPS geometry (rotated dipole) is appropriate to perform the EP-LP designed in section IV.C.

Thus, the proposed circuit has been used to design a new two layer EP-LP converter, which includes the materials where the dipoles must be printed. A 0.5 mm-thick GML 1000 substrate ($\varepsilon_r$=3.38, tan$\delta$=0.003) has been chosen as supporting material for both layers, and a period of P=7 mm is considered (W=0.5 mm). Thus, the equivalent circuit used to perform the design is shown in Fig. 20a. The final dimensions obtained after the design process are: ($L_1$=5 mm and $\alpha_1$=-60.5), ($L_2$=5 mm and $\alpha_2$=-88.5°) and d=2.5 mm (Fig. 15c). Note that the lengths of the dipoles and the air spacer are different from those obtained in section IV.C, mainly because of the presence of the dielectrics. Note also that the thickness (d=2.5 mm) is made up of two 0.5 mm-thick supporting dielectrics and a 1.5 mm-thick air spacer. The circuit elements in accordance with Fig. 9 and Fig. 20a of the final dimensions are: ($Z_{d1}$)=($C_a$=1.109 fF, $L_a$=13.838 nH, $C_b$=0.897 fF, $L_b$=12.454 nH, $C_c$=-2.575 fF, $L_c$=-0.355 nH) and ($Z_{d2}$)=($C_a$=0.365 fF, $L_a$=4.010 nH, $C_b$=-22.009 fF, $L_b$=-0.002 nH, $C_c$=7.781 fF, $L_c$=5.129 nH).

Since the thicknesses of the substrates are fixed, the electrical performance of this new design is slightly worse than that obtained in section IV.C at 20 GHz: XP<-25 dB, reflection levels better than ($R_1$=-18 dB, $R_2$=-24 dB) and insertion loss lower than 0.1 dB.

The designed EP-LP converter has been manufactured using micromachining techniques, and the resulting prototype is made up of 24x24 cells. A 1.5 mm-thick framework of dielectric material was also manufactured to perform the air spacer.

Fig. 20 shows the manufactured device placed in the test system used to measure the S parameters. The measurement system is that conventionally used to measure PPS-based devices in transmission or reflection under periodical conditions, which consists of a pair of rectangular horns that work as transmitter/receiver. The position between the horns and the DUT (Device Under Test) was selected to obtain an appropriate illumination level of -25 dB at the edges of the device whilst the center of the device is excited by a locally plane field with the appropriate angle of incidence. To characterize the transmission parameters, the horns are placed at both sides of the DUT, so that the incident field impinges normally on the device. However, the two horns must be placed at the same side of the DUT to characterize the reflection parameters. Thus, a tilt angle of 10° must be assumed to appropriately place the two horns in the specular direction to each other and to avoid coupling.

Fig. 21 shows the measured reflections and transmissions (as defined in section IV.A) of the manufactured EP-LP converter as a function of the frequency. As can be seen, the device converts the elliptical incident polarization into a purely linear polarization at 21 GHz (XP level better than -30 dB), and provides reflections of (R1=-10 dB and R2=-18 dB) and an insertion loss lower than 0.3 dB. Although this design was performed to operate at 20 GHz, a displacement of the spectral response can be appreciated, which could be attributed to manufacturing tolerances or testing errors (note that the measurements of the S parameters in reflection were done at the angle 10°). However, the measurements validate the circuit and the viability of the cell based on a rotated dipole to design EP-LP or EP-CP converters. Fig. 21 also shows the simulations of the device using both the equivalent circuit of Fig. 20a and CST, which includes the effect of the losses (not accounted in the equivalent circuit). It can be appreciated that the circuit is able to simulate the multilayer structure of Fig. 20a with a good accuracy. In this case, although high order modes are propagating in the dielectrics (the cutoff frequency is 17 GHz) and the separations between discontinuities are thinner than $\lambda_0/7$, the vacuum spacers make these modes are appropriately attenuated, as discussed in section II.D.

## V. CONCLUSION

A Foster's equivalent circuit of a Planar Periodic Structure (PPS) with an arbitrary geometry is presented. The proposed bi-mode circuit (4-port network) shows an invariant circuit topology to the PPS geometry, is completely made up of invariant-frequency lumped elements and is the simplest in terms of number of elements. The topology makes it possible to carry out a standard process to obtain the circuit independently of the geometry of the scatter and allows using the classical circuit theory to design a multitude of devices. The circuit is suitable within the bi-mode bandwidth of the cell for normal incidence and when the dielectric spacers are thicker than $\lambda/8$.

In the particular cases of a defected slotted ring or a rotated dipole, six lumped elements are required to accurately predict the electrical behavior within the whole bi-mode bandwidth of the cell for any arbitrary set of dimensions.

The cell based on the rotated dipole has been also evaluated to design EP-LP or EP-CP converters with the aid of the equivalent circuit, which is demonstrated as an efficient pre-design tool. In the case of devices based on a single layer, the incident elliptical polarization must exhibit a relatively low axial ratio (AR<2) to obtain acceptable return loss. If the axial ratio is higher than 2, two layers are necessary at least to achieve reasonable XP levels (<30 dB) and the cancelation of the reflections. The bandwidth of the device depends on the number of layers, so that a narrow bandwidth is achieved using a two-layer structure.